\documentclass{aastex631}

\usepackage{multirow}

\begin{document}

\title{Long-term radio monitoring of the fast X-ray transient EP240315a: evidence for a relativistic jet}

\author[0000-0003-4631-1528]{Roberto Ricci}
\affiliation{Dipartimento di Fisica,
Universit\'a di Tor Vergata,
Via della Ricerca Scientifica, 1, 
00133 Rome, Italy}
\affiliation{INAF-Istituto di Radioastronomia,
Via Gobetti, 101,
40129, Bologna, Italy}

\author{Eleonora Troja}
\affiliation{Dipartimento di Fisica 
Universit\'a di Tor Vergata 
Via della Ricerca Scientifica, 1, 
00133 Rome, Italy}

\author[0000-0003-0691-6688]{Yu-Han Yang}
\affiliation{Dipartimento di Fisica 
Universit\'a di Tor Vergata 
Via della Ricerca Scientifica, 1, 
00133 Rome, Italy}

\author[0009-0004-9520-5822]{Muskan Yadav}
\affiliation{Dipartimento di Fisica 
Universit\'a di Tor Vergata 
Via della Ricerca Scientifica, 1, 
00133 Rome, Italy}

\author[0009-0007-6104-337X]{Yuan Liu}
\affiliation{National Astronomical Observatories, 
Chinese Academy of Sciences,
100101 Beijing, China}

\author[0000-0002-9615-1481]{Hui Sun}
\affiliation{National Astronomical Observatories, 
Chinese Academy of Sciences, 
100101 Beijing, China}

\author[0000-0002-6299-1263]{Xuefeng Wu}
\affiliation{Purple Mountain Observatory, 
Chinese Academy of Sciences, 
210023 Nanjing, China}

\author[0000-0003-2516-6288]{He Gao}
\affiliation{Institute for Frontier in Astronomy and Astrophysics, 
Beijing Normal University, 
100875 Beijing, China}
\affiliation{School of Physics and Astronomy,
Beijing Normal University,
100875 Beijing, China}

\author[0000-0002-9725-2524]{Bing Zhang}
\affiliation{Nevada Center for Astrophysics and Department of Physics and Astronomy, University of Nevada Las Vegas,
Las Vegas, NV 89154, USA}

\author{Weimin Yuan}
\affiliation{National Astronomical Observatories, 
Chinese Academy of Sciences, 
100101 Beijing, China}
\affiliation{School of Astronomy and Space Sciences, 
University of Chinese Academy of Sciences, 
Beijing, 100049, China}

%\collaboration{20}{(AAS Journals Data Editors)}

%\author{F.X Timmes}
%\affiliation{Arizona State University}
%\affiliation{AAS Journals Associate Editor-in-Chief}

%\altaffiliation{AASTeX v6+ programmer}

%\affiliation{AAS Director of Publishing}
%\affiliation{American Astronomical Society \\
%1667 K Street NW, Suite 800 \\



\begin{abstract} 
The recent launch of Einstein Probe (EP) in early 2024 opened up a new window onto the transient X-ray sky, allowing for real-time discovery and follow-up of fast X-ray transients (FXRTs). Multi-wavelength observations of FXRTs and their counterparts are key to characterize the properties of their outflows and, ultimately, identify their progenitors. Here, we report our long-term radio monitoring of EP240315A, a long-lasting ($\sim 1000$ s) high redshift ($z=4.9$) FXRT associated to GRB~240315C. Our campaign,  carried out with the Australian Telescope Compact Array (ATCA), followed the transient's evolution at two different frequencies (5.5 GHz and 9~GHz) for three months. 
In the radio lightcurves we identify an unusual steep rise at 9 GHz, possibly due to a refreshed reverse shock, and a late-time rapid decay of the radio flux, which we interpret as a jet break due to the outflow collimation. 
We find that the multi-wavelength counterpart of EP240315A is well described by a model of relativistic jet seen close to its axis, with jet half-opening angle $\theta_j \approx 3 ^{\circ}$ and beaming-corrected total energy $E \simeq 4\times 10^{51}$~erg, typical of GRBs. These results show that a substantial fraction of FXRTs may be associated to standard GRBs and that sensitive X-ray monitors, such as Einstein Probe and the proposed HiZ-GUNDAM and Theseus missions, can successfully pinpoint their relativistic outflows up to high-redshifts. 

\end{abstract}



\section{Introduction} \label{sec:intro}

The core-collapse of the most massive stars is known to produce
short-lived flashes of high-energy radiation, known as gamma-ray bursts (GRBs; \citealt{WoosleyBloom2006}), traditionally discovered using wide field of view (FoV) gamma-ray and hard X-ray monitors. 
%GRBs are considered unique probes of star-formation across cosmic times \citep{LambReichart2000, Greiner2012, Lien2014}, however at the highest redshifts the bulk of their emission moves to lower energies in the observer's frame. 

Since their discovery in the early 70s \citep{Klebesadel73}, our physical understanding of these explosions has been based on their observed properties in the gamma-ray band \citep[e.g.][]{Kouveliotou93}.
In particular, the duration of the prompt gamma-ray emission 
is widely used to guide the identification of the GRB progenitors and to constrain the lifetime of the GRB central engine \citep[e.g.][]{KumarZhang2015, Petropoulou2020}. 

The discovery of X-ray flares \citep{Burrows2005}, sudden rebrightenings of low-energy ($<$10 keV) radiation that follow many GRBs, suggested instead that the central engine could remain active on much longer timescales than the observed GRB duration \citep{FanWei2005, Zhang2006, Liang2006, Lazzati2007, MaxhamZhang2009}.
Although multiple models were proposed to explain X-ray flares, 
ranging from delayed MHD instabilities \citep{Giannios2006} to structured jets \citep{Duque2022}, those requiring a prolonged activity of the engine better reproduce their broadband properties \citep{Falcone2007,Troja2015,Yi2016}. 
This aspect was recently demonstrated by the X-ray transient discovered by the Einstein Probe (EP) mission on March 15, 2024, and dubbed EP240315A \citep{GCNEP}. 
At a redshift $4.9$ \citep{GCNredshift}, EP240315A is among the most distant high-energy transients ever known. Its temporal and spatial coincidence with GRB240315C \citep{GCNGRBs}, as well as the large energy output implied by its distance scale \citep[$E_{\gamma,\rm iso}\approx 6\times10^{53}$ erg,][]{EPpaper}, identify this event as a GRB. 
%Using cosmological parameters from \citep{Planck2020} 

A notable feature of EP240315A is the striking difference between its high-energy and low-energy time profiles \citep{EPpaper}. 
At energies above 15 keV, observed by \textit{Swift}/BAT and Konus-Wind, it displays a multi-peaked light curve with a duration of approximately 40 s (observer's frame),
which is fairly typical for GRBs. 
At lower (0.3-4 keV) energies, observed by the Wide X-ray Telescope (WXT) aboard EP, the outburst is seen to last over 1,000~s,  which is among the longest GRB durations ever measured. Even when correcting for time dilation effects, an intrinsic duration of $\approx$200 s remains among the top 1.5\% of GRB durations \citep{Minaev2020}. 
Some of the observed X-ray peaks occur after the prompt gamma-ray episode, and are thus similar to X-ray flares.
Some other X-ray pulses are detected even before the gamma-ray signal, setting back the time of the explosion by at least six minutes \citep{EPpaper}.

Observations of EP240315A naturally link X-ray flares to the prompt gamma-ray emission as part of the same outburst from a long-lived central engine. 
Although the gamma-ray phase remains dominant in terms of energetics, this case shows that the standard gamma-ray window might greatly underestimate the lifetime of the GRB engine, both before and after the main prompt emission episode. 
This has fundamental implications for our understanding of these extreme explosions: powering such long-lasting activity is non-trivial for standard GRB central engines, including accreting black holes (BHs) and magnetars \citep{Perna2006,Dai2006,Proga2006}. 
Studies of X-ray flares suggest that this phenomenon may affect over 30\% of the GRB population \citep{Chincarini2007,Yi2016}.
The occurrence of soft X-ray emission preceding the prompt gamma-rays by hundreds of seconds is instead less constrained by observations. 
Only additional cases of joint X-ray/gamma-ray detections would clarify whether this behavior is common or is a peculiar trait of EP240315A. 

In this paper, we explore whether EP240315A/GRB240315C is representative of the general GRB population by studying the properties of its relativistic outflow constrained through multi-wavelength afterglow observations. 
Whereas the GRB afterglow faded fast in the 
X-ray and optical band, it remained visible for weeks at radio wavelengths, allowing us to detect an achromatic steepening of its flux. 
We interpret this feature as a jet-break \citep{Rhoads1999,Sari1999,vanEerten2012} and use it to constrain the geometry and total energy release of the GRB outflow.

The paper is organized as follows: in Section~\ref{sec:obs} we describe the multiwavelength observations, focusing on the ATCA monitoring campaign; in Section~\ref{sec:afmod} we detail the afterglow modelling first in empirical terms and then following a physical parametrization; in Section~\ref{sec:conc} we discuss the implications of our results.

We adopt flat cosmology with parameters $H_0 = 67.4$  km s$^{-1}$ Mpc$^{-1}$ and $\Omega _m = 0.315$ \citep{Planck2020}, which gives the luminosity distance of $D_L = 46.1$ Gpc.

\section{Observations} \label{sec:obs}

\subsection{Australian Telescope Compact Array}
\label{sec:radio} % used for referring to this section from elsewhere

\begin{figure*}[!t]
\centering
\vspace{0.5cm}
\includegraphics[angle=0,scale=0.65]{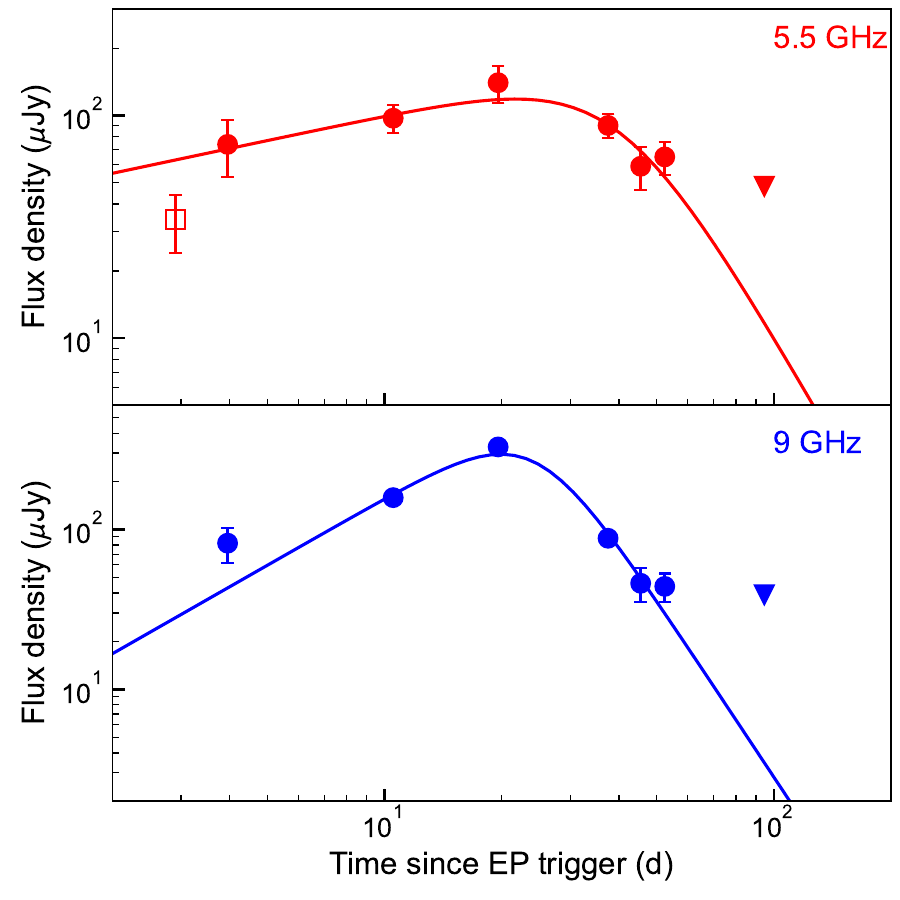}
\caption{Smoothed double power law fit to the light curve of the 
5.5 GHz (upper panel) and 9~GHz (lower panel) radio data. Downward triangles represent 3\,$\sigma$ upper limits from ATCA. The open square shows the MeerKAT data point, shifted from 3 GHz to 5.5 GHz.} 
\label{fig:sdplfit}
\end{figure*}

EP240315A was observed with the ATCA on seven separate runs as part of a 
Target-of-Opportunity program CX564 between March 19th and June 18th 2024. The radio interferometer was in array configuration 6A and later 6D (the maximum extension). The long hiatus between the 6th and 7th Epoch was due to a change in the array configuration (H168) between May 16th and June 4th, which was unfavourable for the target detection at high resolution. All observing runs were carried out in continuum mode with 1-MHz channel correlator mode in C/X-band at the standard centre frequencies of 5.5 and 9 GHz, with a bandwidth of 2 GHz each. The primary and bandpass calibrators were 1934$-$638 and 0823$-$500, and the complex gains calibrator was 0941$-$080. The data were reduced in Miriad \citep{sault1995} with standard procedures for flagging, calibration and imaging. A robustness parameter value r=0 was used at the Fourier inversion stage as a good trade-off between natural and uniform weighting in order to minimize the map rms noise in source detection experiments.

All images were Gaussian fitted and the peak flux densities values were extracted at the transient source position, being the best estimate of flux density in the case of point-like source observations.  The rms noise in the maps were evaluated in areas near the transient source position but far away from bright sources. As the target was quite faint in all images the multiplicative term (5\% of the peak flux density, accounting for the residual complex gain error) turned out to be negligible in computing the flux density uncertainty budget. 
The resulting measurements are reported in Table~\ref{tab:radioflux}, which lists the elapsed time since the event trigger, the flux density and corresponding 1\,$\sigma$ error bar at 5.5 GHz and at 9.0 GHz.

\subsection{Other radio facilities}

In the radio band, EP240315A was also observed by e-MERLIN \citep{Bruni2024, Rhodes2024} and by MeerKAT \citep{Carotenuto2024}. 
The e-MERLIN observations resulted in $5\,\sigma$ upper limits of 75 $\mu$Jy at 5.5~d and 105 $\mu$Jy at 12.5~d, both at a frequency of 5 GHz. 
These measurements are lower than the fluxes derived with ATCA at a similar frequency and epoch (Table~\ref{tab:radioflux}), and point to a day-to-day variability of the observed radio flux. Strong flux modulations are not uncommon in GRB radio light curves and are often ascribed to the effects of interstellar scintillation (ISS; \citealt{GranotvanderHorstREVIEW}).
To estimate these effects we use the NE2001 model of the Galactic distribution of free electrons \citep{CordesLazio2002}. From this we derive a scattering measure SM\,=\,$3.11\times 10^{-4}$\,kpc\,m$^{-20/3}$ and a critical frequency $\nu_0$\,=\,11.3~GHz below which ISS might be relevant.

At lower frequencies, MeerKAT reported a weak detection of the radio counterpart at 
2.9~d with a flux density of $\approx30~\mu$Jy at 3 GHz in a map with an rms noise of 8.5 $\mu$Jy/beam. The measured flux is lower by a factor of $\approx$2 than the simple extrapolation of the ATCA spectrum. This is consistent with ISS effects although, as discussed below, could also be an intrinsic feature of the afterglow spectrum due to self-absorption.

\subsection{X-ray and optical follow-up}

 X-ray observations of EP240315A were performed with the Follow-up X-ray Telescope (FXT) on board EP, starting about 42 hr after the trigger, and with the Chandra X-ray Observatory, starting about 72 hr after the trigger \citep{EPpaper}. A fading X-ray afterglow was detected 
at a flux of 1.8$_{-0.7}^{+1.4} \times 10^{-13}$ erg s$^{-1}$cm$^{-2}$ (90\% confidence level; c.l.). and remained detectable for approximately 8 days after the trigger. 

 At optical and near-infrared (nIR) wavelengths, the counterpart was detected by multiple facilities for about 16 days, as reported by \citet{EPpaper}.  
%To the dataset presented by \citet{EPpaper}, we add the photometry of the PRime-focus Infrared Microlensing Experiment (PRIME) \textbf{TBD}. 
Magnitude corrections for Galactic extinction were implemented using a reddening of \(E(B-V) = 0.05\) mag \citep{Schlegel1998} 
and the extinction law described by \cite{Cardelli1989} with $R_V \approx$3.1.

\begin{table}
	\centering
	\caption{ATCA radio flux densities in the EP240315a monitoring campaign. Col. 1:
 elapsed time since WXT trigger T0 in days; cols. 2-3: flux density and its error bar at 5.5~GHz; cols. 4-5: flux density at 9~GHz with its error bar.  Upper limits are reported at 3\,$\sigma$ level.}
	\label{tab:radioflux}
	\begin{tabular}{lcccc} 
		\hline
		  T$-$T0 & S$_{\rm 5.5GHz}$ & $\pm$ & S$_{\rm 9GHz}$ & $\pm$  \\
             (d) &  ($\mu$Jy)  & ($\mu$Jy) & ($\mu$Jy) & ($\mu$Jy)  \\
		\hline
		  3.95   &    74       &         21   &   82      &        20      \\
           10.5	 &    97       &         14   &  158      &        12      \\
           19.6	 &   140       &         27   &  328      &        31      \\
           37.5   &    90       &         11   &   88      &         8      \\
           45.4   &    59       &         13   &   46      &        11      \\
	   52.4  &    65      &          11  &    44      &         9      \\ 
           94.5  &  $<$ 48   &          $-$  &   $<$ 39   &        $-$      \\
%           94.5  &    3      &          16  &     7      &        13      \\
           \hline
	\end{tabular}
\end{table}

\section{Afterglow Modeling} \label{sec:afmod}

\subsection{Basic properties}
The GRB afterglow is a broadband synchrotron radiation produced by a population of shock-accelerated electrons, whose energy distribution can be described as a simple power-law function with index $p$. 
At any time, the afterglow spectral shape consists of smoothly joint power-law segments \citep{Sari1998,GranotSari2002},
fully characterized by four quantities: the cooling frequency $\nu_c$, the characteristic synchrotron frequency  $\nu_m$, the self-absorption frequency $\nu_a$, and the peak flux $F_{\rm pk}$. 

Preliminary constraints on these parameters can be derived from the observed afterglow evolution \citep[e.g.][]{WijersGalama1999} by assuming that the main emission mechanism is the standard forward shock (FS), arising from the interaction between the relativistic outflow and its surrounding interstellar material (ISM) \citep{ReesMeszaros1992}. 
When possible, we describe the monochromatic afterglow flux as a power-law in both frequency and time, $F_{\nu} \propto \nu^{-\beta} t^{-\alpha}$. For the radio data, 
a simple power-law cannot reproduce the long-term temporal evolution, and we adopt a smoothly broken power-law \citep{Beuermann1999}. 

First, we consider the optical and X-ray data, which follow a similar temporal decay with slope $\alpha=$1.66$\pm$0.05. 
This indicates that they lie on the same segment of the synchrotron spectrum with slope $\beta_{OX} = (2 \alpha +1) / 3  \approx 1.5$ for $\max(\nu _m, \nu_c) \lesssim \nu_O<\nu_X$ during the time frame of observations 1~d$\lesssim t \lesssim$10 d. 
Within the standard GRB scenario, the resulting electrons' spectral index is rather steep, $p$\,$\approx$\,2$\beta_{OX}$\,$\approx$\,3. 

Once established that the optical and X-rays belong to the same segment, we can infer the amount of intrinsic extinction from the spectral energy distribution. According to the SED around 2 d after the burst, including $H$-, $z$-, $I$-band and X-ray data, we derive ${E(B-V)}_z<0.011$ ($3\sigma$ c.l.) for the extinction law described by \cite{Cardelli1989}. 

Then, we turn to the radio afterglow which is seen to rise with a slope of $\alpha_r = -0.4^{+0.3}_{-0.2}$ at 5.5 GHz, consistent with the spectral segment $\nu_a < \nu_r < \nu_m$.  The rising trend favors a circumburst medium with uniform density since in a medium with a wind-like density profile the expected radio light curve is flat. 
At the higher frequency (9 GHz), the observed slope of $-1.4^{+0.4}_{-0.3}$ is steeper than predictions of a simple FS model and may include some contributions from a fast rising cooling reverse shock (RS). 
To test this interpretation, we model the 9 GHz light curve with two broken power-law models: the former tracks the FS emission and is characterized by the same parameters of the 5.5 GHz curve, the latter tracks the possible RS and its parameters are free to vary. 
This fit constraints the rising RS slope to $\alpha_{\mathrm{RS}}$ steeper than $-$2.5, which, considering also the lack of RS features at lower frequencies, 
is only consistent with standard closure relations \citep[e.g.][]{Gao2013} 
for $\nu_{a, \mathrm{RS}} \gtrsim 9$~GHz (observer's frame). 
However, this scenario faces some problems. For instance, the collision that occurred at such late time should have taken place far away from the central engine, when both the blast wave and the density of the jet ejected later would have been very low. In this case, it would be difficult to achieve such large $\nu_{a, \mathrm{RS}}$.

At $t\approx40$~d the radio emission is seen to steeply decline at both frequencies with slope $3.1^{+1.0}_{-1.2}$ and $3.7^{+0.6}_{-0.7}$, respectively. 
Very few mechanisms can account for such large change in slope ($\Delta \alpha_r \gtrsim$3) and rapid decline. For instance, 
a factor of ten drop in the density of the external medium would induce  at most $\Delta \alpha_r \approx$0.5 \citep[e.g.][]{Gat2013}, whereas the passage of the peak frequency $\nu_m$ would cause $\Delta \alpha_r \approx$2 for $p\approx3$ \citep{Sari1998}. 
A bright RS crossing the 9 GHz band could produce a steep decay \citep[e.g.][]{Dichiara2022}, although not as fast as the observed value. 

A late-time drop in flux can only be explained with a jet-break, that occurs when the jet edges become visible to the observer \citep{Rhoads1999, Sari1999}. 
We consider synchrotron radiation from the FS as the main mechanism 
powering the observed radiation at 5 GHz. Within this scenario, the post-break flux is expected to decline as  $\propto t^{-1/3}$ if $\nu_r < \nu_m$
and as $\propto t^{-p}$ if $\nu_r > \nu_m$. 
As we observe a value that is in between the two predictions, the peak frequency $\nu_m$ is likely passing through the observed radio frequencies between 20 d and 40 d. 

Assuming a jet-break occurs at $t_j\gtrsim 20 / (1+z) \approx 3.3$\,d, we can infer the jet opening angle \citep[e.g.][]{Sari1999,vanEerten2012}:

\begin{equation} \label{eq:thj}
    \theta_j \approx 3.6^{\circ} \left( \frac{t_j}{\rm 3.5~d} \right)^{3/8}
    \left( \frac{E_{\rm K,iso}}{\rm 10^{54} erg} \right)^{-1/8} \left( \frac{n}{\rm 0.001~cm^{-3}} \right)^{1/8}
\end{equation}
where $E_{\rm K,iso}$ is the isotropic-equivalent kinetic energy of the blast-wave, and $n$ the density of the environment. 
The narrow collimation of the jet implies that the true gamma-ray energy release is $E_{\gamma}\approx 10^{51}$\, erg, in line with the population of GRBs \citep[e.g.][]{Wang2018}. 

\subsection{Energy Injection}
Between 250 and 450 ks ($\approx$13 hr rest-frame) 
a rebrightening of the optical and X-ray light curves suggests that EP240315A experienced a late episode of energy injection. 
The bump displays a small amplitude and a smooth profile with a short rise time ($\Delta\,t/t \sim 0.5$) and shallower decay ($\Delta\,t/t \sim 1$), which does not necessarily require a reactivation of the central engine. 
Various other mechanisms can reproduce similar small bumps in an afterglow light curve, such as density jumps in the external medium, 
complex jet structures, and refreshed shocks arising from the collision between fast-moving and slow-moving ejecta \citep{Rees1999,Granot2003}. 
Since the complex temporal profile and strong spectral evolution of the prompt emission point to multiple episodes of ejections from the central engine, we focus on the latter model.
A refreshed shock would also induce a reverse shock back into the ejecta, 
which could explain the steep rise of the radio light curve at 9 GHz. 

To estimate the bulk Lorentz factor of the fast moving ejecta, we 
use the time of the first afterglow detection as an upper limit to the deceleration time $t_d$ \citep{Piran1999,zhang2018}: 
\begin{equation} 
    \Gamma_0 > 76  \left( \frac{t_d}{\rm 0.1~d} \right)^{-3/8}
    \left( \frac{E_{\rm K,iso}}{\rm 10^{54} erg} \right)^{1/8} \left( \frac{n}{\rm 0.1~cm^{-3}} \right)^{-1/8}
\end{equation}

Using a simple internal shock model \citep{Moss2023}, the average Lorentz factor of the slow moving ejecta can then be estimated as
$\Gamma \approx 10 (t / 13 {\rm hr})^{3/8} (n_{-2}/E_{52})^{1/8}$ , where  $n_{-2}$ is the density of the circumburst medium in units of $10^{-2}$ cm$^{-2}$, and  $E_{52}$ is the isotropic equivalent energy
of the fast ejecta in units of $10^{52}$ erg. 
To reproduce the relatively fast rise of the bump, a narrow spread in Lorentz factors is also required, $\Delta \Gamma / \Gamma \approx 0.2$.

\subsection{Broadband modeling}

 We model the GRB afterglow using the open-source Python package \texttt{afterglowpy}, based on the semi-analytic models of \citet{vanEerten2010} and \citet{Ryan2020}. 
In its simplest version, the afterglow is fully described by five free parameters: isotropic kinetic energy $E_{\rm K, iso}$, density of the ISM $n$, distribution index of electrons $p$, the %shock microphysical parameters 
fractions of shock energy into electrons $\epsilon_e$ and magnetic field $\epsilon_B$. The beaming of the outflow adds an additional parameter, half-opening angle $\theta_j$.
The fraction of shock-accelerated electrons $\xi_N$ is a poorly constrained parameter, and is generally fixed to 1 to avoid degeneracy. In our fit we explore three different cases: $\xi_N=1$ for comparison with the majority of GRB afterglow fits, $\xi_N=0.1$ motivated by particle-in-cell simulations of relativistic shocks, and $\xi_N$ free to vary.

\begin{table}
	\centering
	\caption{Best-fit parameters from the afterglow model fitting. Case 1: $\xi_N = 1$. Case 2: $\xi_N =0.1$. Case 3: $\xi_N$ is a free parameter. We fix $\theta_{\rm obs} = 0$ for all the cases. BIC is the Bayesian Information Criterion \citep{Schwarz1978} for the best-fit results: the model with lower BIC is the preferred one.}
	\label{tab:agres}
	\begin{tabular}{ccccc}
		\hline
Parameter  &  Prior &   \multicolumn{3}{c}{Posterior}    \\  
        &            &      Case 1      &       Case 2     &        Case 3    \\   
\hline  
% \multirow{7}{}{Parameter} 
  log $E_{k,\mathrm{iso}}$ (erg)& (50, 60)    & $53.3^{+ 0.5}_{-0.4}$ & $54.4^{+ 0.6}_{-0.4}$ &  $56.7^{+ 0.9}_{-1.4}$            \\
%theta_obs     &        (0,pi/2)   &                  &                  &                    \\ 
    log $\theta_j$ (rad)   &      (-3, -0.5)  &        $-1.2^{+0.2}_{-0.2}$            &        $-1.3^{+ 0.2}_{-0.3}$            &      $-1.6^{+ 0.3}_{-0.3}$                \\
   log $n$ (cm$^{-3}$)     &      (-6, 2)    &       $-2.3^{+ 1.3}_{-1.4}$           &         $-2.5^{+ 1.3}_{-1.6}$         &     $-2.4^{+1.9}_{-1.6}$               \\
     p                &      (2, 3.5)    &      $2.9^{+ 0.2}_{-0.06}$            &            $2.9^{+ 0.2}_{-0.06}$      &    $2.9^{+ 0.2}_{-0.05}$                \\
    log $\epsilon_e$ &     (-6, -0.5)   &           $-1.2^{+ 0.3}_{-0.7}$       &      $-2.4^{+ 0.3}_{-0.3}$            &    $-4.7^{+1.4}_{-0.9}$                \\
   log $\epsilon_B$ &     (-6, -0.5)   &          $-0.9^{+ 0.4}_{-0.8}$        &   $-0.9^{+ 0.4}_{-0.8}$               &          $-1.6^{+0.8}_{-1.4}$          \\
        log $\xi_N$      &       (-5, 0)    &       0 (fixed)           &         -1 (fixed)         &           $-2.9^{+1.2}_{-0.8}$         \\
\hline
% \multirow{2}{}{Statistics}
  $\chi ^2$/dof & $-$     &  42.81/32   &  42.12/32  &   41.28/31   \\
        BIC              &        $-$     &   64.6   &   63.9   & 66.7\\
\hline
	\end{tabular}
\end{table}

\begin{figure}
    \centering
    \includegraphics[width=75mm]{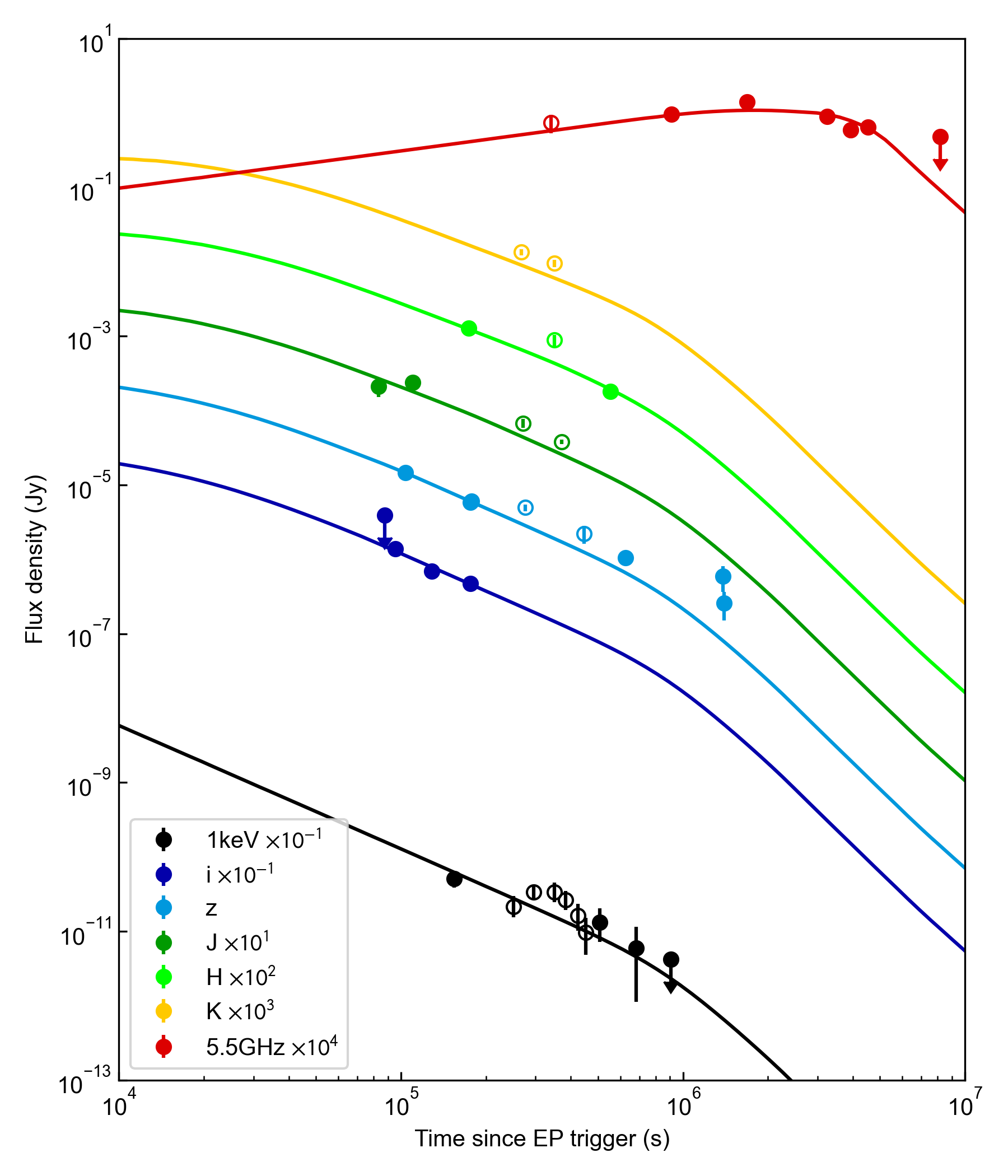}
    \includegraphics[width=85mm]{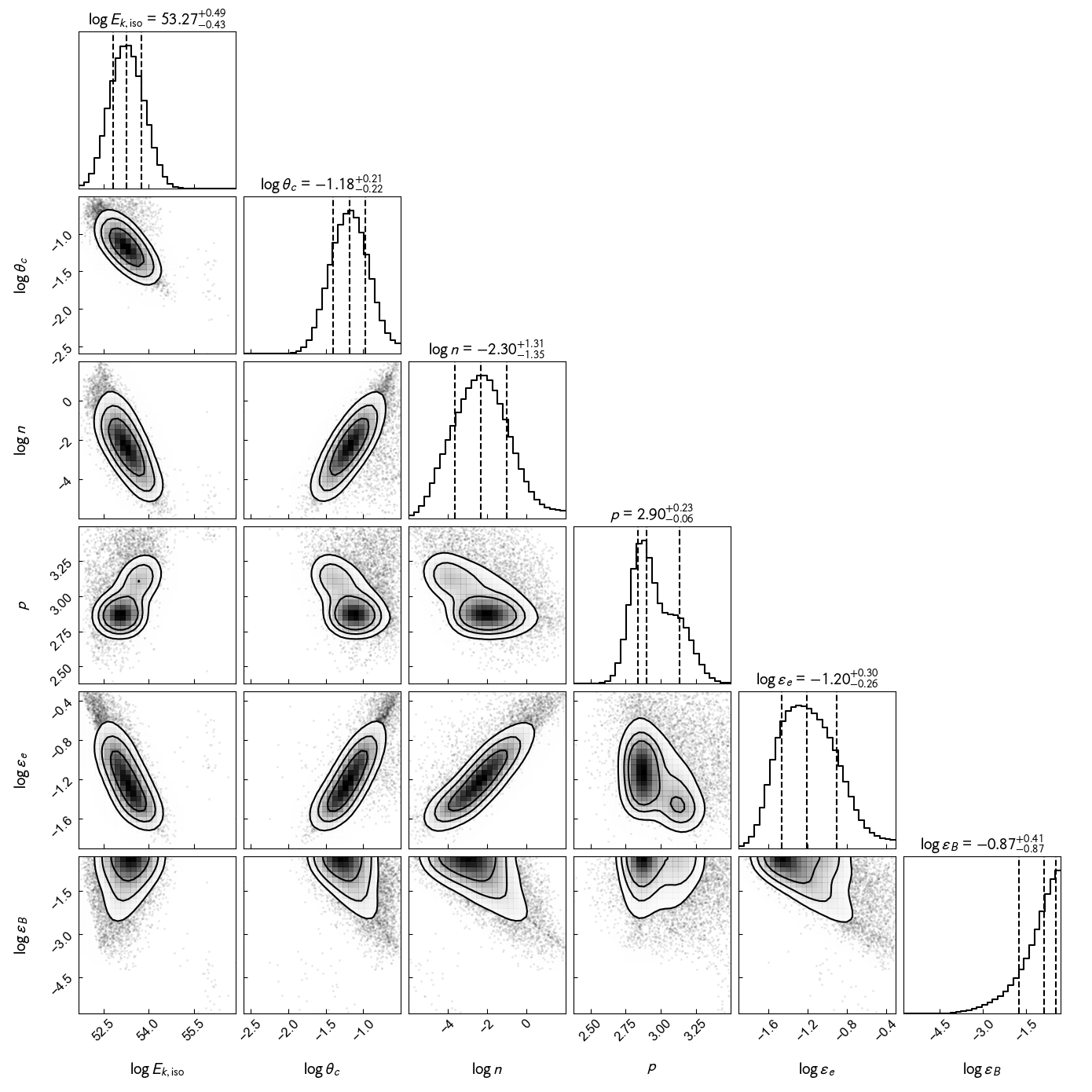}     
    \caption{Broadband modelling results. Left panel: best-fit light curves with observing points from the radio to the X-ray band. Downward arrows represent upper limits. Empty points are used for observations in the exclusion time span (250-450 ks after trigger time). Right panel: corner plot of the posterior distribution of the modelled parameters. In both panels Case~1 is presented.}
    \label{fig:agres}
\end{figure}

Additionally, the code allows the user to account for viewing angle effects through the parameter $\theta_{\rm obs}$ ($\theta_{\rm obs}$=0 corresponds to an on-axis observer), and to explore multiple angular structures of the GRB jet (e.g. uniform, Gaussian, and power-law). 
In our fits, we explore the simple standard case of a uniform jet viewed on-axis, which is already sufficient to provide a good description of the dataset.

The code does not include the RS emission, which may contribute above 9 GHz, and the synchrotron self-absorption regime, which may be relevant below 3 GHz.  For these reasons, we only include the observations at 5.5 GHz to model the FS afterglow. 
Energy injection is included in \texttt{afterglowpy} but with limited options, thus we exclude from our fit the X-ray and optical data between 250 ks and 450 ks. The results of the afterglow model fitting are summarized in Table~\ref{tab:agres} for the three cases explored: $\xi_N =1$ (Case~1), $\xi_N = 0.1$ (Case~2) and $\xi_N$ free to vary (Case~3). The model fit light curves from the radio to the X-ray band and posterior parameter distributions are shown in Fig.~\ref{fig:agres} for the Case~1 only. 

The fit identifies a jet-break occurring at around 20 d after the GRB, which translates into a jet opening angle $\theta_j \approx 3.6^{\circ}$ (Case 1) and $2.9^{\circ}$ (Case 2). 
The evidence of a jet-break is mainly driven by the radio data, as it occurs after the end of the X-ray and optical observations. 
The two fits with fixed $\xi_N$ yields similar values of $\theta_j$ and are equivalent in terms of reduced $\chi^2$ and BIC values. 
However, Case 1 with $\xi_N$=1 yields an isotropic-equivalent kinetic energy that is three times lower than the value measured in gamma-rays, implying a high gamma-ray efficiency $\eta_{\gamma} \gtrsim$ 70\%.  
This would be challenging for most prompt emission mechanisms and, in particular, for internal shocks in hydrodynamic jets \citep{Daigne1998,BeniaminiPiran2013}.
Instead, Case 2 ($\xi_N$=0.1) eases the requirement on the gamma-ray production with a dissipative efficiency of $\eta_{\gamma} \sim$ 20\%. 
By using the best fit values of $E_{k,\mathrm{iso}}$ and $\theta_j$,  we infer a beaming-corrected kinetic energy $E_K$ = $ (1 - cos\ \theta_j) E_{k,\mathrm{iso}} \approx  4\times 10^{50}$ erg (Case 1) and $3 \times 10^{51}$ erg (Case 2) , in keeping with cosmological GRBs \citep{Wang2018}. 
By leaving $\xi_N$ free to vary (Case 3), the fit does not substantially improve and the total kinetic energy tends to unphysically high values. 

All our fits tend to prefer a tenuous environment with a uniform density between $10^{-4}$ and $10^{-2}$, which is consistent with the interstellar medium. Using these best fit values, we infer a self-absorption frequency $\nu_{\mathrm{a}} \approx 1$ GHz (observer's frame), close to the range covered by MeerKAT. Finally, it is worth noting the soft value of the electron's spectral index $p\approx 3$. Basic shock acceleration principles set this value to 2.23 \citep{KeshetWaxman2005}, although observations of GRB afterglows display slightly softer spectra with an average $p\approx 2.4$ \citep{Curran2006}. \citet{Warren2017} found that if acceleration is efficient, then nonlinear effects linking the shock structure to the accelerated particles lead to a harder electron's spectrum. Our soft $p$ value is not consistent with these findings and may point instead to non-efficient particle acceleration. 

\begin{figure}
    \centering
    \includegraphics[width=150mm]{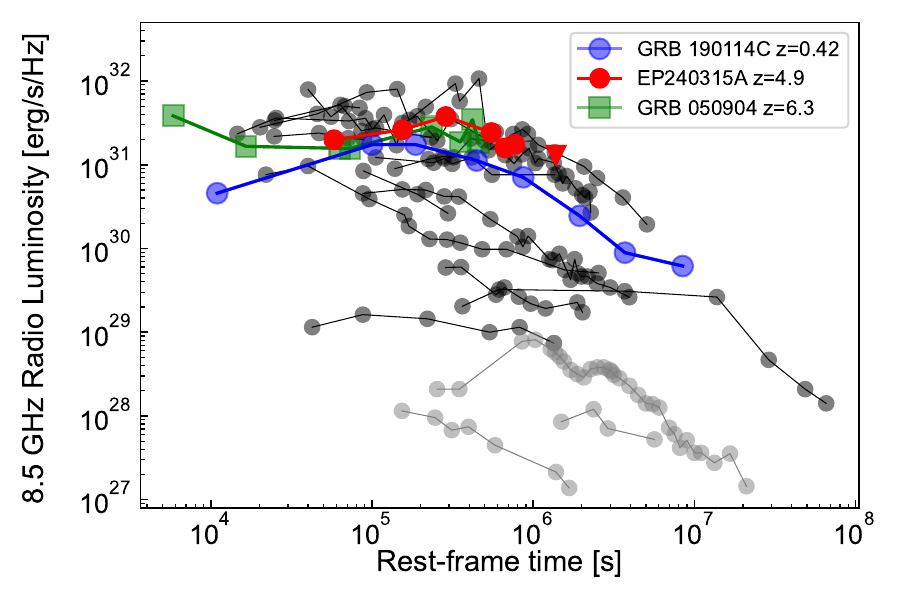}    
    \caption{Radio luminosity light curve of EP240315A compared with a sample of cosmological GRBs (black) and sub-energetic GRBs (gray) \citep{Dichiara2022,ChandraFrail2012}. }
    \label{fig:lum}
\end{figure}

In Figure~\ref{fig:lum} we present a comparison between the radio afterglow of EP240315A and a representative sample of GRBs. As derived from our broadband modeling, EP240315A displays properties fully consistent with the population of cosmological GRBs. 
Whereas its prompt emission is relatively faint in the soft X-ray band \citep{EPpaper}, its radio counterpart occupies the brightest tail of the afterglow distribution, with a luminosity comparable to GRB~050904 \citep{ChandraFrail2012} and higher than GRB190114C \citep{MagicPaper}. 

\section{Conclusions} \label{sec:conc}

At the time of writing, EP has publicly reported 18 FXRTs. 
Of these, only two were detected by \textit{Swift}/BAT, including EP240315A, whereas most of the other ones happened outside the BAT's field of view (FoV). 
Additional constraints come from \textit{Fermi}/GBM observations: 
only two possible gamma-ray counterparts were reported, although in 13 cases the FXRT happened during normal operations and had good visibility ($\theta_{\rm NaI}<60$\,deg). This suggests that a substantial fraction of the FXRT population ($\gtrsim$25\%) may overlap with GRBs. 

We used the case of EP240315A to characterize the relativistic outflow powering some of these FXRTs and compare it to the properties of cosmological GRBs. Our study finds that EP240315A was produced by a collimated relativistic jet with a narrow opening angle $\theta_j \approx 3^{\circ}$ and total energy 
$E \approx  4\times 10^{51}$ erg, in keeping with the population of long GRBs. Whereas in the case of EP240315A, its temporal and spatial coincidence with GRB240315C unambiguously classifies it as GRB, our study shows that similar relativistic explosions can be discovered  by sensitive X-ray surveys and successfully identified through the evolution of their multiwavelength counterparts. 
Although more events are needed to better understand the interplay of GRBs and FXRTs, this case confirms that missions such as Einstein Probe and the proposed hiZ-GUNDAM \citep{Yonetoku2020} and THESEUS \citep{Amati2021} will help us pinpoint the explosions of the most massive stars up to the highest redshifts.

\begin{acknowledgments}
We thank Simone Dichiara for sharing the data on GRB radio afterglows.
ET, RR, YY and MY acknowledge the support of the European Research Council through the Consolidator grant BHianca (Grant agreement ID: 101002761). 
The Australia Telescope Compact Array is part of the Australia Telescope National Facility (https://ror.org/05qajvd42) which is funded by the Australian Government for operation as a National Facility managed by CSIRO. We acknowledge the Gomeroi people as the Traditional Owners of the Observatory site. 
\end{acknowledgments}

\bibliography{references}{}
\bibliographystyle{aasjournal}



\end{document}